\documentclass[twocolumn,english,aps,prb,twocolum,superscriptaddress,bibnotes,amsmath,amssymb,floatfix]{revtex4-2}
\usepackage[colorlinks=true,citecolor=blue,linkcolor=magenta]{hyperref}

\usepackage[markup=nocolor, authormarkupposition=left]{changes} 
\usepackage{soul}
\usepackage[utf8]{inputenc}
\usepackage[english]{babel}
\usepackage{amsmath,amsfonts,amssymb}
\usepackage[T1]{fontenc}
\usepackage{url}

\usepackage{amsmath}
\usepackage{amsfonts}
\usepackage{amssymb}

\usepackage{epstopdf}
\usepackage{graphicx}
\graphicspath{{./Figures/}}

\usepackage{changes}

\begin{document}

\title{Universal Kerr-thermal dynamics of self-injection-locked microresonator dark pulses}

\author{Shichang Li}
\thanks{These authors contributed equally to this work.}
\affiliation{Shenzhen Institute for Quantum Science and Engineering, Southern University of Science and Technology, Shenzhen 518055, China}
\affiliation{International Quantum Academy, Shenzhen 518048, China}

\author{Kunpeng Yu}
\thanks{These authors contributed equally to this work.}
\affiliation{International Quantum Academy, Shenzhen 518048, China}
\affiliation{Hefei National Laboratory, University of Science and Technology of China, Hefei 230088, China}

\author{Dmitry A. Chermoshentsev}
\thanks{These authors contributed equally to this work.}
\affiliation{Russian Quantum Center, Moscow 143026, Russia}

\author{Wei Sun}
\email[]{sunwei@iqasz.cn}
\affiliation{International Quantum Academy, Shenzhen 518048, China}

\author{Jinbao Long}
\affiliation{International Quantum Academy, Shenzhen 518048, China}

\author{Xiaoying Yan}
\affiliation{Shenzhen Institute for Quantum Science and Engineering, Southern University of Science and Technology, Shenzhen 518055, China}
\affiliation{International Quantum Academy, Shenzhen 518048, China}

\author{Chen Shen}
\affiliation{International Quantum Academy, Shenzhen 518048, China}
\affiliation{Qaleido Photonics, Shenzhen 518048, China}

\author{Artem E. Shitikov}
\affiliation{Russian Quantum Center, Moscow 143026, Russia}

\author{Nikita Yu. Dmitriev}
\affiliation{Russian Quantum Center, Moscow 143026, Russia}

\author{Igor A. Bilenko}
\affiliation{Russian Quantum Center, Moscow 143026, Russia}
\affiliation{Faculty of Physics, Lomonosov Moscow State University, Moscow 119991, Russia}

\author{Junqiu Liu}
\email[]{liujq@iqasz.cn}
\affiliation{International Quantum Academy, Shenzhen 518048, China}
\affiliation{Hefei National Laboratory, University of Science and Technology of China, Hefei 230088, China}

\maketitle

\noindent\textbf{Microcombs, formed in optical microresonators driven by continuous-wave lasers, are miniaturized optical frequency combs. 
Leveraging integrated photonics and laser self-injection locking (SIL), compact microcombs can be constructed via hybrid integration of a semiconductor laser with a chip-based microresonator.  
While the current linear SIL theory has successfully addressed the linear coupling between the laser cavity and the external microresonator, it fails to describe the complicated nonlinear processes, especially for dark-pulse microcomb formation. 
Here, we investigate -- theoretically, numerically and experimentally -- the Kerr-thermal dynamics of a semiconductor laser self-injection-locked to an integrated silicon nitride microresonator. 
We unveil intriguing yet universal dark-pulse formation and switching behaviour with discrete steps, and establish a theoretical model scrutinizing the synergy of laser-microresonator mutual coupling, Kerr nonlinearity and photo-thermal effect. 
Numerical simulation confirms the experimental result and identifies the origins. 
Exploiting this unique phenomenon, we showcase an application on low-noise photonic microwave generation with phase noise purified by 23.5 dB. 
Our study not only adds critical insight of pulse formation in laser-microresonator hybrid systems, but also enables all-passive, photonic-chip-based microwave oscillators with high spectral purity. 
}

Optical frequency combs (OFCs) \cite{Udem:02, Cundiff:03, Fortier:19, Diddams:20} have revolutionized timing, spectroscopy, precision measurement, and testing fundamental physics. 
Microcombs -- harnessing enhanced Kerr nonlinearity in optical microresonators driven by continuous-wave (CW) pumps -- have allowed miniaturized OFCs with small size, weight and power consumption \cite{DelHaye:07,Kippenberg:18}. 
With the emergence and quick maturing of ultralow-loss photonic integrated circuits \cite{Moss:13, Gaeta:19, Zhang:17, Liu:21}, as well as heterogeneous and hybrid integration with semiconductor lasers \cite{Stern:18, Xiang:21}, today microcombs can be built entirely on-chip and manufactured in large volume with low cost, catalyzing wide deployment outside laboratories. 

Depending on the microresonator's group velocity dispersion (GVD), there are two types of coherent microcombs. 
The bright dissipative soliton microcombs \cite{Herr:14, Yi:15, Brasch:15, Liang:15, Joshi:16, He:19} require anomalous GVD, while the dark-pulse (also termed ``platicon'') microcombs require normal GVD \cite{Xue:15, Lobanov:15, Huang:15b, Parra-Rivas:16, Nazemosadat:21, WangH:22}.
Compared to solitons, platicons exhibit remarkably higher CW-to-pulse power conversion efficiency \cite{Xue:17b, JangJ:21}, thus are advantageous for coherent optical communication \cite{Fulop:18} and photonic microwave generation \cite{Sun:25}.  
To initiate platicons, laser self-injection locking (SIL) \cite{Kondratiev:17, Kondratiev:23} offers the most robust and effective form, and permits seamless integration of the pump laser and the microresonator on a monolithic substrate \cite{Voloshin:21, Jin:21, Lihachev:22a}. 
However, while the current \textit{linear} SIL theory and model \cite{Kondratiev:17} have addressed the linear coupling between the laser cavity and the external microresonator, it fails to manage the complicated \textit{nonlinear} processes in platicon formation.

\begin{figure}[t!]
\centering
\includegraphics{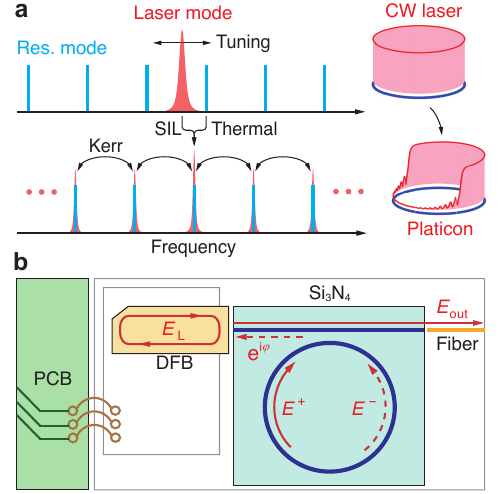}
\caption{
\textbf{Principle and schematic of SIL, Kerr nonlinearity and photo-thermal effect in the laser-microresonator coupled system. }
\textbf{a}. 
Frequency- and time-domain picture describing the synergy of SIL, photo-thermal effect and Kerr nonlinearity, which yields narrowing of the pump laser's linewidth, red-shift of the microresonator's resonance grid, and formation of a platicon microcomb. 
Res. mode: resonance mode.
\textbf{b}. 
Experimental setup.
A DFB laser chip edge-coupled to a Si$_3$N$_4$ microresonator chip. 
A printed circuit board (PCB) provides stable current to the laser and stabilizes the laser temperature. 
Output light from the Si$_3$N$_4$ microresonator is collected by a lensed fiber. 
$E_{\mathrm{L}}$: light amplitude in the laser cavity.
$E^{+}$/$E^{-}$: light amplitude in the clockwise/counter-clockwise direction in the microresonator. 
$E_{\mathrm{out}}$: light amplitude at the microresonator output.
$\varphi$: feedback phase of $E^{-}$.
}
\label{Fig:1}
\end{figure}

Here, we investigate the Kerr-thermal dynamics of platicon formation using SIL of a semiconductor laser to an integrated silicon nitride (Si$_3$N$_4$) microresonator.  
We unveil an intriguing platicon switching behaviour with discrete steps, allowing operation of platicons in quiet states immune to laser noise.   
Figure~\ref{Fig:1}a illustrates the principle of our study.
When a laser-cavity mode is tuned into a resonance mode of the Si$_3$N$_4$ microresonator and light is coupled into the microresonator, SIL occurs that locks the laser frequency to the resonance \cite{Kondratiev:17,Kondratiev:23} and suppresses the laser linewidth. 
The linewidth suppression ratio is proportional to $Q_\mathrm{r}^2/Q_\mathrm{d}^2$, where $Q_\mathrm{r/d}$ is the quality factor of the microresonator/laser cavity. 
Meanwhile, photo-thermal effect \cite{Carmon:04, Brasch:16, Yi:16b, Gao:22} induces a global frequency shift of the microresonator's resonance grid. 
With sufficient intracavity power, Kerr nonlinearity induces four-wave mixing (FWM) \cite{Kippenberg:04} that translates photons to other resonances.  
In microresonators of normal GVD, the synergy of SIL and Kerr effect ultimately yield platicon formation. 

\noindent \textbf{Experimental result}. 
We use a commercial DFB laser chip edge-coupled to a Si$_3$N$_4$ microresonator chip \cite{Ye:23,Sun:25}, as shown in Fig.~\ref{Fig:1}b.
The DFB laser operates at 1549 nm.
It has 1 nm wavelength tuning range and 159 mW output CW power with driving current up to 500 mA. 
The Si$_3$N$_4$ microresonator has 10.7 GHz free spectral range (FSR) and normal GVD ($D_2 <0$).
The microresonator is critically coupled \cite{Pfeiffer:17}, with intrinsic quality factor $Q_0= 23\times 10^6$.  
Actual photographs and characterization data of the DFB lasers and the Si$_3$N$_4$ microresonators are found in Supplementary Information Note 1.

Light $E_\mathrm{L}$ in the laser cavity is emitted and coupled into the microresonator's clockwise direction $E^+$.
Rayleigh scattering in the microresonator reflects some light to the counter-clockwise direction $E^-$ and to the laser cavity. 
To facilitate SIL, we optimize the edge coupling and the gap distance (thus the feedback phase $\varphi$) between the two chips. 
In the SIL regime, we continuously tune the laser current, and monitor the output optical power and frequency from the Si$_3$N$_4$ chip. 
The output light beats against a reference CW laser, and the beat frequency is recorded. 
The forward (backward) tuning corresponds to increasing (decreasing) laser current, which decreases (increases) laser frequency. 

\begin{figure*}[t!]
\centering
\includegraphics{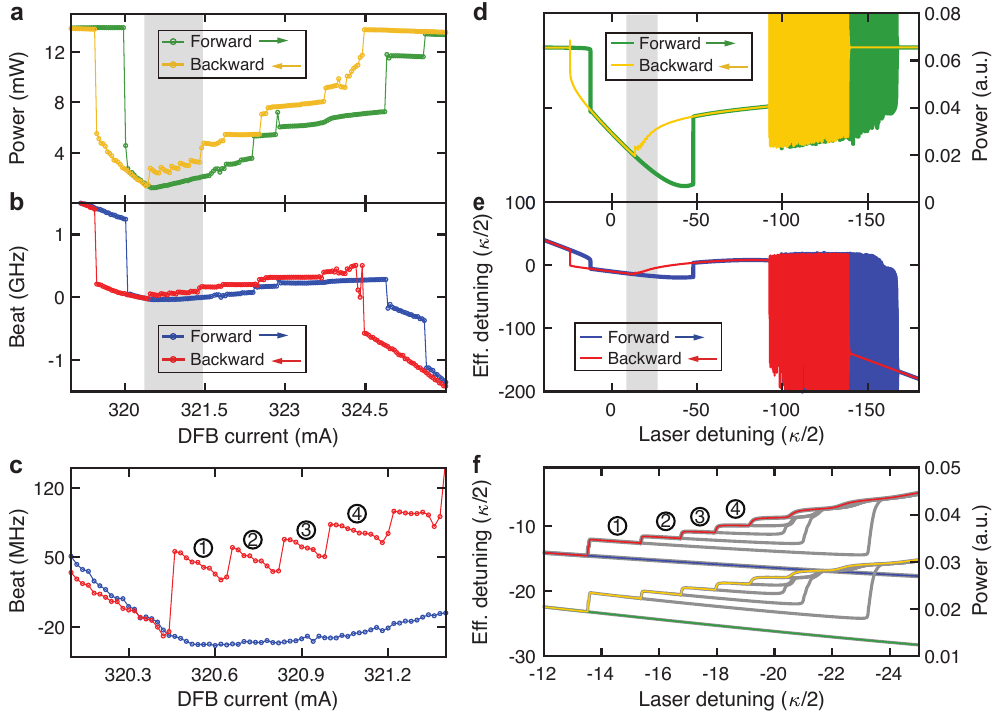}
\caption{
\textbf{Experimental result in comparison with numerical simulation}.
\textbf{a}.     
Experimentally measured output laser power with forward (green) or backward (yellow) tuning.
\textbf{b}. 
Experimentally measured beat frequency between the output laser and a frequency-fixed reference laser, with forward (blue) or backward (red) tuning.
\textbf{c}.
Zoom-in profiles of the gray-shaded region in \textbf{b} with forward (blue) or backward (red) tuning.
Discrete steps with backward tuning are numbered with 1 to 4. 
\textbf{d}, \textbf{e}.
Numerical simulation results corresponding to experimental data in \textbf{a}, \textbf{b}.
Horizontal axes are the frequency detuning (in the unit of $\kappa /2$) between the free-running laser frequency and the cold resonance frequency.
Effective (Eff.) detuning is the frequency detuning (in the unit of $\kappa /2$) between the laser frequency at the microresonator output and the cold resonance frequency. 
\textbf{f}.
Zoom-in profile of the gray-shaded zoom in \textbf{d}, \textbf{e}.
Gray curves outline the full detuning range of platicon steps.
}
\label{Fig:2}
\end{figure*}

Figure~\ref{Fig:2}(a, b) shows the experimental data.
When SIL occurs, the output power experiences a sudden drop, and the output frequency is shifted.
In the SIL regime, we observe synchronous dynamics of the output power and the frequency, due to the \textbf{photo-thermal effect} \cite{Carmon:04, Brasch:16, Yi:16b, Gao:22}.
For example, the decreasing output power corresponds to decreasing transmitted power and increasing intracavity power. 
The latter causes an increasing resonance red-shift (towards longer wavelength) via the photo-thermal effect\cite{Carmon:04}.
As the laser frequency is locked to the resonance, the red-shifted resonance drags the laser frequency, leading to decreasing output frequency.

\begin{figure*}[t!]
\centering
\includegraphics{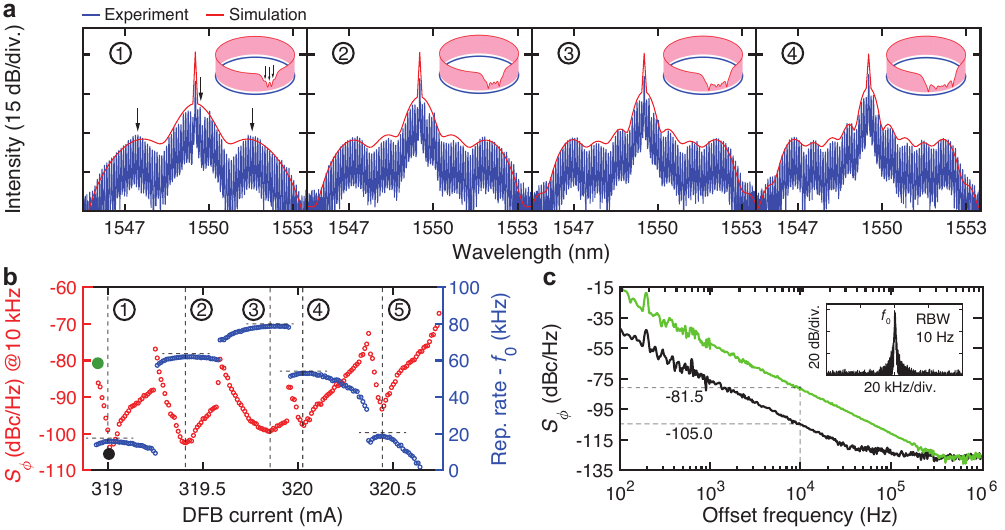}
\caption{
\textbf{Optical spectra of different platicon states and the resulted noise-quenching effect}.
\textbf{a}.
Experimentally measured (blue lines) and simulated (red curves) optical spectra of the platicon states labeled in Fig. \ref{Fig:2}c, f.  
They agree not only on the spectral envelopes but also on the number of fringes (marked with arrows). 
Insets: simulated time-domain pulse shapes of the corresponding platicon states. 
The fringes, marked with arrows, are related to the oscillating tails at the bottom of the dark pulse. 
\textbf{b}. 
Measured platicon's repetition rate $f_\mathrm{rep}$ (blue) and phase noise $S_{\phi}$ at 10 kHz Fourier offset frequency (red) of the 10.7-GHz microwave, with backward tuning. 
$f_0 =10.68545$ GHz.
Vertical dashed lines highlight that the local minimum of $S_{\phi}$ coincides with the local maximum of $f_\mathrm{rep}$. 
Horizontal dashed lines highlight that this coincidence is associated with $\mathrm{d}f_\mathrm{rep}/\mathrm{d}I=0$. 
\textbf{c}. 
Phase noise $S_{\phi}$ spectra of the local maximum and minimum points marked with green and black dots in \textbf{b}. 
Phase noise quenching up to 23.5 dB is observed.  
Inset shows the beat note of the lowest $S_{\phi}$ (black data) with 10 Hz RBW.  
}
\label{Fig:3}
\end{figure*}

Besides, it is apparent that the output power and frequency exhibit discrete step features.  
This is contrary to the conventional linear SIL model \cite{Kondratiev:17} showing continuous, nearly anchored tuning curves of the frequency. 
Here, the appearance of these steps is attributed to the \textbf{Kerr nonlinearity} in the microresonator with sufficient intracavity power. 
Note that similar discontinuous curves have been observed and characterized in self-injection-locked solitons \cite{Voloshin:21}, while here we observe and characterize these curves for the first time for platicons.   
Figure~\ref{Fig:2}c shows the zoom-in profile of the gray-shaded zone in Fig. \ref{Fig:2}b. 
Steps in backward tuning with 320.4 to 321.3 mA laser current are marked with 1 to 4.
Typical optical spectra within each step are shown in Fig.~\ref{Fig:3}a, evidencing formation and switching of different platicon states. 
Distinct fringes are marked with arrows on the spectral envelopes.
Numerical simulation of time-domain pulse shapes in Fig.~\ref{Fig:3}a (discussed later) confirms that all these platicon states are ``single platicons'' comprising only one dark pulse in the microresonator. 

Moreover, identical Kerr-thermal platicon dynamics has also been observed in other two independent SIL setups, where different DFB lasers and a 21.3-GHz-FSR Si$_3$N$_4$ microresonator are used. 
Details are found in Supplementary Information Note 1 and 2. 
These parallel experiments suggest that, our observation of the Kerr-thermal dynamics and discrete steps is universal and independent of the particular devices.

\vspace{0.2cm}
\noindent\textbf{Theory and numerical simulation}. 
To better understand our experimental result, we establish a comprehensive theoretical model synergizing SIL, Kerr nonlinearity, and photo-thermal effect.
The equations describing our laser-microresonator nonlinear system are written as
\begin{widetext}
\begin{align}
\partial_{\tau}N &= \gamma_\mathrm{e}\left[ \gamma-F(N-1)-(N-1+\gamma)|a_\mathrm{L}|^2\right] \\
\partial_\tau a_\mathrm{L} &=\frac{1}{2}\gamma_\mathrm{p}\left (1+i\alpha_\mathrm{g}\right )(N-1)a_\mathrm{L}+i\mathrm{e}^{i(\alpha_\mathrm{L}\tau+\varphi)}\frac{\kappa_\mathrm{c}}{f_\mathrm{e}}b_0 \\
\partial_\tau a_{\mu} &=-\left(1+i\sum_{n=2}^{\infty}\frac{d_{n}}{n!}\mu^{n}-iT\right)a_{\mu}+i\left(\mathcal{F}[|a|^{2}a]_{\mu}+2a_{\mu}\sum_{\mu^{\prime}}\left|b_{\mu^{\prime}}\right|^{2}\right)+i\tilde{\beta}b_{\mu}+i\mathrm{e}^{i(-\alpha_\mathrm{L}\tau+\varphi)}f_\mathrm{e} a_\mathrm{L}\delta_{0\mu} \\
\partial_{\tau}b_{\mu} &=-\left(1+i\sum_{n=2}^{\infty}\frac{d_{n}}{n!}\mu^{n}-iT\right)b_{\mu}+i\left(\mathcal{F}[|b|^{2}b]_{\mu}+2b_{\mu}\sum_{\mu^{\prime}}\left|a_{\mu^{\prime}}\right|^{2}\right)+i\tilde{\beta}a_{\mu} \\
\partial_{\tau}T &=\left[ k_{T}\sum_{\mu}\left(\left|a_{\mu}\right|^{2}+\left|b_{\mu}\right|^{2}\right)-T\right ]/t_\mathrm{h} .
\end{align}
\end{widetext}
Derivation of Eq. (1--5) is found in Supplementary Information Note 3.
In short, Eq. (1) describes the time evolution of the electron carrier density $N$ normalized to the threshold carrier density inside the laser cavity, 
with $\tau = \kappa t/2$,
$\gamma_\mathrm{e} = [2I/(\kappa e V_\mathrm{L}N_0)-2\kappa_{N}/\kappa]/\gamma$,
$\gamma = G_0/(G_\mathrm{n}N_0)$, 
$F = 2\kappa_{N}/\kappa/\gamma_\mathrm{e}$. 
Here, $\kappa/2\pi$ is the loaded linewidth of the Si$_3$N$_4$ microresonator, 
$t$ is the real time, 
$I$ is the driving current on the laser diode,
$e$ is the electron charge, 
$V_\mathrm{L}$ is the effective volume of the laser diode's active area, 
$\kappa_N$ is the dissipation rate of carrier density,
$N_0$ is the electron carrier density of the free-running laser with a fixed driving current,
$G_0$ is the gain of the light intensity per second at $N_0$ in the laser cavity,
$G_\mathrm{n}$ is the gain coefficient per second,
$a_\mathrm{L}$ represents the light amplitude in the laser cavity.

Equation (2) describes the laser dynamics in the laser cavity, 
with $\gamma_\mathrm{p} = 2G_\mathrm{n} N_0/\kappa$,
$\alpha_\mathrm{L}=2\delta \omega_\mathrm{L} / \kappa$,
$\varphi = \omega_0 t_\mathrm{s}$, 
$\kappa_\mathrm{c} = 4\kappa_\mathrm{ex} T_\mathrm{L} \sqrt{T_\mathrm{cL}T_\mathrm{cR}} / (\kappa^2 \tau_\mathrm{L})$, 
$f_\mathrm{e} = 2 \sqrt{\gamma_\mathrm{e} \kappa_\mathrm{ex} T_\mathrm{L} T_\mathrm{cR} g V_\mathrm{L}}/\sqrt{\kappa^2 G_\mathrm{n} \tau_\mathrm{L}}$.
Here, 
$\delta \omega_\mathrm{L}$ represents the frequency difference between the cold-cavity resonance mode $\omega_\mathrm{L}$ and the microresonator resonance mode $\omega_0$,
$t_\mathrm{s}$ is the time delay between the laser cavity and the microresonator, 
$\kappa_\mathrm{ex}$ is the external coupling rate of the microresonator, 
$T_\mathrm{L}$ is the light transmission efficiency at the laser cavity's output surface, 
$T_\mathrm{cR}$ is the light coupling efficiency from the laser cavity to the microresonator,
$T_\mathrm{cL}$ is the light coupling efficiency from outside into the laser cavity, 
$\tau_\mathrm{L}$ is the light round-trip time in the laser cavity,
$g$ is the nonlinear coefficient of the Si$_3$N$_4$ microresonator.
$\alpha_\mathrm{g}$ is the linewidth enhancement factor and $b_0$ represents the back-scattered light in the microresonator at the $0^\mathrm{th}$ resonance mode.

Equations (3, 4) describe the dynamics of the forward-propagating light $a_\mu$ and the back-scattered light $b_\mu$ at the $\mu ^\mathrm{th}$ resonance mode,  
with $d_n = 2D_n / \kappa$,
and $\tilde{\beta} = 2\beta / \kappa$.
Here, $D_n$ is the $n^\mathrm{th}$-order dispersion of the microresonator, 
$\beta$ is the back-scattering ratio between the back-scattered light and the forward-propagating light, 
and $\mathcal{F}[|a|^{2}a]$ ($\mathcal{F}[|b|^{2}b])$ represents the Kerr interaction.

Equation (5) describes the evolution of the normalized frequency shift $T$ induced by the temperature in the microresonator, 
with $k_T = K_T/g$, 
and $t_\mathrm{h} = \kappa \tau_\mathrm{h}/2$. 
Here, $K_T$ is the thermal-induced resonance shift coefficient, 
and $\tau_\mathrm{h}$ is the thermal relaxation time.

We further perform numerical simulation of Eqs. (1--5) using experimentally obtained or realistic values for each parameter.   
Details are found in Supplementary Information Note 3.
Figure~\ref{Fig:2}(d,e) presents the simulated output optical power and frequency, which agree with the experimental data on both the overall trend (due to the photo-thermal effect) and the platicon steps (due to the Kerr nonlinearity).
Figure~\ref{Fig:2}f shows the zoom-in profile of the gray-shaded zone in Fig.~\ref{Fig:2}(d,e), and highlights the full detuning range of platicon steps with gray curves.
Steps corresponding to those in Fig.~\ref{Fig:2}c are marked. 
We emphasize that, previous efforts \cite{Kondratiev:20, Lihachev:22a, WangH:22} to model platicon formation in the SIL regime fail to investigate and analyze the transient behaviour, and thus fail to reveal the step features. 
Meanwhile, they have also omitted the photo-thermal effect. 
Although the Lugiato-Lever equation (LLE) can capture platicon steps,it does not contain the SIL dynamics.
Detailed comparison between our model and the LLE model is found in Supplementary Information Note 4.

In Fig.\ref{Fig:3}a, the red curves presents the simulated platicon spectra for each labeled steps in Fig.~\ref{Fig:2}f. 
They confirm the experimental data, not only on the spectral envelopes but also on the number of fringes (marked with arrows). 
Figure~\ref{Fig:3}a insets present the simulated time-domain pulse shapes, revealing that the fringes are related to the oscillating tails (marked with arrows). 
Unlike the steps of bright dissipative solitons \cite{Guo:16, Voloshin:21}, we demonstrate here that platicon steps can occur even in the single-platicon state, i.e. only one dark pulse in the microresonator. 

\vspace{0.2cm}
\noindent\textbf{Low-noise photonic microwave generation}.
Finally, we showcase an immediate application of our findings, i.e. to generate low-noise photonic microwave. 
Photonic microwave -- microwave synthesized via photonics -- allows unrivalled noise performance and bandwidth breaking the bottleneck of their electronic counterparts \cite{Fortier:11, Li:14, Xie:17}. 
Microcombs offer an appealing solution for low-noise microwave and terahertz generation \cite{Liu:20, ZhangS:19a, Wang:21, Tetsumoto:21, Yao:22}.  
Photodetection of the microcomb's pulse stream creates a microwave whose carrier frequency corresponds to the microcomb's repetition rate $f_\mathrm{rep}$. 
Several methods have been implemented to further improve the microwave's spectral purity, including the use of an external microwave \cite{Weng:19}, an auxiliary laser \cite{LiuR:24}, a transfer comb \cite{Lucas:20}, or two-point optical frequency division \cite{Kudelin:24, SunS:24, Zhao:24}. 

Here we harness the platicon switching dynamics, and demonstrate a unique effect allowing suppression of the microwave's phase noise. 
In contrast to the methods mentioned above, our system is free-running, all passive (i.e. without any active locking), and simple.
Experimentally, we collect the output light from the microresonator with a commercial photodetector (PD).  
The PD converts the dark pulse stream of $f_\mathrm{rep}=10.7$ GHz to a 10.7-GHz microwave in the critical X-band. 

By varying the laser current $I$ within the SIL regime, we monitor the platicon's $f_\mathrm{rep}$ and phase noise $S_{\phi}$ at 10 kHz Fourier offset frequency of the 10.7-GHz microwave, as shown in Fig. \ref{Fig:3}b. 
For each step, a local minimum of $S_{\phi}$ is observed. 
Meanwhile, it coincides with the local maximum of $f_\mathrm{rep}$, as highlighted with vertical dashed lines in Fig. \ref{Fig:3}b. 
This coincidence is due to that, at the local maximum of $f_\mathrm{rep}$, we have $\mathrm{d}f_\mathrm{rep}/\mathrm{d}I=0$, as highlighted with horizontal dashed lines in Fig. \ref{Fig:3}b. 
Therefore,  $f_\mathrm{rep}$ becomes insensitive to current-noise-induced laser frequency jitter, resulting in the lowest $S_{\phi}$. 
More analysis of this coincidence is found in Supplementary Information Note 5. 
In fact, this effect is similar to the ``quiet point'' effect observed in bright dissipative solitons without SIL \cite{Yi:17,Triscari:23}. 
The underlying mechanism is attributed to the multi-mode coupling between the laser cavity and the Si$_3$N$_4$ microresonator, which causes asymmetric comb-line enhancement or suppression, as shown in Fig.~\ref{Fig:3}a.
As a result, $f_\mathrm{rep}$ depends on the laser frequency (be more specifically, the detuning of the free-running laser frequency to the cold resonance frequency). 
In our theoretical model, only the pump resonance is considered coupled, i.e. $b_0$ in Eq. (2).
A more precise simulation on the optical spectra should include multi-mode coupling by substituting $b_0$ with $\sum_{\mu} b_\mu \mathrm{e}^{i\mu d_1 (\tau -\tau_\mathrm{s})}$.

The $S_{\phi}$ spectra of the local maximum and minimum points, marked with green and black dots in Fig.~\ref{Fig:3}b, are measured and compared in Fig.~\ref{Fig:3}c, showing 23.5 dB noise reduction. 
The lowest $S_{\phi}$ reaches $-45/-75/-105$ dBc/Hz at 0.1/1/10 kHz Fourier offset frequency, and is limited by the shot noise floor at higher offset frequency\cite{Savchenkov:08a}.
Figure~\ref{Fig:3}c inset shows the beat note of the lowest $S_{\phi}$ (black data) with 10 Hz resolution bandwidth (RBW). 
Supplementary Information Note 6 illustrates that, the identical phenomenon has been also observed in parallel experiments with different Si$_3$N$_4$ microresonators and DFB lasers, and thus is universal. 

\vspace{0.2cm}
In conclusion, we unveil an intriguing yet universal Kerr-thermal dynamics of a semiconductor laser self-injection-locked to a Si$_3$N$_4$ microresonator where platicon microcombs are formed. 
We experimentally observe and characterize the platicon switching dynamics with discrete steps. 
We further establish a comprehensive theoretical model cooperating the laser-cavity dynamics, the platicon formation dynamics, the photo-thermal dynamics, and the mutual coupling between them.
Numerical simulation confirms the experimental result, and illuminates that the intriguing rich platicon switching dynamics is originated from the synergy of SIL, Kerr nonlinearity and the photo-thermal effect. 
Exploiting this finding, we showcase low-noise platicon-based microwave generation, and achieve 23.5 dB phase noise reduction of the 10.7-GHz microwave carrier.
Our study not only adds critical insight into pulse formation in linear-and-nonlinear-coupled laser-microresonator systems, but also offers a neat solution for photonic-chip-based microwave oscillators with high spectral purity, ideal for microwave photonics, coherent optical communication, analog-to-digital conversion, wireless links, and radar. 

\medskip
\begin{footnotesize}

\vspace{0.1cm}

\noindent \textbf{Acknowledgments}: 
We thank Baoqi Shi for characterizing the Si$_3$N$_4$ chips. 
We acknowledge support from the National Natural Science Foundation of China (Grant No.12261131503, 12404436),
Innovation Program for Quantum Science and Technology (2023ZD0301500), 
National Key R\&D Program of China (Grant No. 2024YFA1409300),
Guangdong-Hong Kong Technology Cooperation Funding Scheme (Grant No. 2024A0505040008), 
Shenzhen-Hong Kong Cooperation Zone for Technology and Innovation (HZQB-KCZYB2020050), 
and Shenzhen Science and Technology Program (Grant No. RCJC20231211090042078). 
Russian Quantum Center was supported by RSF grant (23-42-00111). 
The DFB lasers were fabricated by Shenzhen PhotonX Technology Co. Ltd. and Henan Shijia Photons Technology Co. Ltd. .
Silicon nitride chips were fabricated by Qaleido Photonics.

\vspace{0.1cm}
\noindent \textbf{Author contributions}: 
W. S. and J. Liu conceived the experiment.
S. L., W. S., J. Long and X. Y. built the experimental setup. 
S. L. and W. S. performed the experiments and analyzed the data, with the assistance from A. E. S. and N. Y. D.. 
K. Y., D. A. C., and W. S. developed the theory and performed the simulation, with the assistance from A. E. S., N. Y. D. and I. A. B..
C. S. fabricated the Si$_3$N$_4$ chips.
W. S., S. L., K. Y. and J. Liu wrote the manuscript, with input from others. 
J. Liu supervised the project.

\vspace{0.1cm}
\noindent \textbf{Data Availability Statement}: 
The code and data used to produce the plots within this work will be released on the repository \texttt{Zenodo} upon publication of this preprint.

\end{footnotesize}
\bibliographystyle{apsrev4-1}
\bibliography{bibliography}
\end{document}


\title{Supplementary Information to: Universal Kerr-thermal dynamics of self-injection-locked microresonator dark pulses}

\author{Shichang Li}
\thanks{These authors contributed equally to this work.}
\affiliation{Shenzhen Institute for Quantum Science and Engineering, Southern University of Science and Technology, Shenzhen 518055, China}
\affiliation{International Quantum Academy, Shenzhen 518048, China}

\author{Kunpeng Yu}
\thanks{These authors contributed equally to this work.}
\affiliation{International Quantum Academy, Shenzhen 518048, China}
\affiliation{Hefei National Laboratory, University of Science and Technology of China, Hefei 230088, China}

\author{Dmitry A. Chermoshentsev}
\thanks{These authors contributed equally to this work.}
\affiliation{Russian Quantum Center, Moscow 143026, Russia}

\author{Wei Sun}
\email[]{sunwei@iqasz.cn}
\affiliation{International Quantum Academy, Shenzhen 518048, China}

\author{Jinbao Long}
\affiliation{International Quantum Academy, Shenzhen 518048, China}

\author{Xiaoying Yan}
\affiliation{Shenzhen Institute for Quantum Science and Engineering, Southern University of Science and Technology, Shenzhen 518055, China}
\affiliation{International Quantum Academy, Shenzhen 518048, China}

\author{Chen Shen}
\affiliation{International Quantum Academy, Shenzhen 518048, China}
\affiliation{Qaleido Photonics, Shenzhen 518048, China}

\author{Artem E. Shitikov}
\affiliation{Russian Quantum Center, Moscow 143026, Russia}

\author{Nikita Yu. Dmitriev}
\affiliation{Russian Quantum Center, Moscow 143026, Russia}

\author{Igor A. Bilenko}
\affiliation{Russian Quantum Center, Moscow 143026, Russia}
\affiliation{Faculty of Physics, Lomonosov Moscow State University, Moscow 119991, Russia}

\author{Junqiu Liu}
\email[]{liujq@iqasz.cn}
\affiliation{International Quantum Academy, Shenzhen 518048, China}
\affiliation{Hefei National Laboratory, University of Science and Technology of China, Hefei 230088, China}

\maketitle


\section{Characterization of DFB lasers and Si$_3$N$_4$ microresonators}
\vspace{0.5cm}
The experimental setups (Setup \#1--3) used to characterize the Kerr-thermal dynamics are shown in Supplementary Fig.~\ref{Fig:S1}a--c.
Setup \#1 is the setup in the main text.
We use two commercial DFB lasers from two different vendors. 
DFB lasers \#1 and \#3 are provided by Henan Shijia Photons Technology, and DFB laser \#2 is from Shenzhen PhotonX Technology.
We use two Si$_3$N$_4$ microresonators (Si$_3$N$_4$ \#1 and \#2) of 10.7 GHz free spectral range (FSR), and one microresonator (Si$_3$N$_4$ \#3) of 21.3 GHz FSR, fabricated by us using the process described in Ref. \cite{Ye:23, Sun:25}.

\begin{figure*}[b!]
\renewcommand{\figurename}{Supplementary Figure}
\centering
\includegraphics{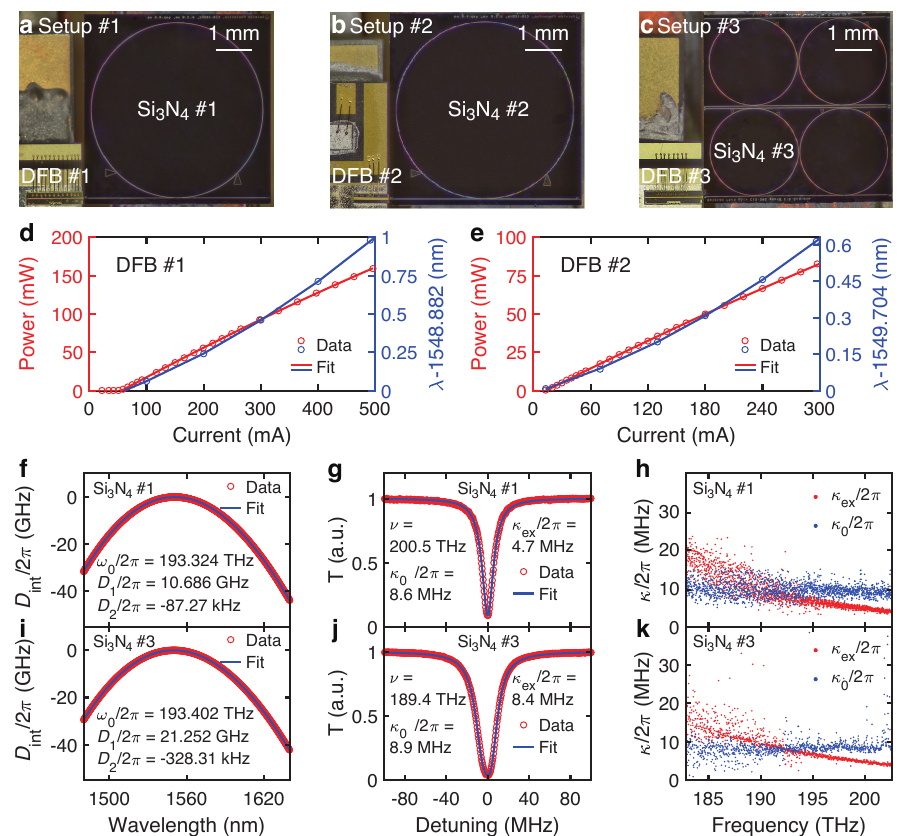}
\caption{
\textbf{Chips characterization of the three SIL setups}.
\textbf{a--c}. 
The photographs of the three different setups (Setup \#1--3).
Si$_3$N$_4$ \#1 and \#2 are the 10.7-GHz-FSR microresonators. 
Si$_3$N$_4$ \#3 is a 21.3-GHz-FSR microresonator. 
DFB \#1 and \#3 are the DFB lasers from  Henan Shijia Photons Technology. 
DFB \#2 is a DFB laser from Shenzhen PhotonX Technology.
The red(blue) circles show the measured output power(wavelength) of DFB laser \#1 in \textbf{d} and DFB laser \#2 in \textbf{e} versus the applied current. 
Solid lines are the corresponding fits.
\textbf{f}(\textbf{i}).
Measured integrated dispersion of Si$_3$N$_4$ microresonator \#1(\#3). 
The microresonator has 10.686(21.252)-GHz FSR and normal GVD of $D_2/2\pi =-87.7(-328.31)$ kHz at the reference frequency $\omega_0/2\pi =193.324(193.402)$ THz.
\textbf{g}(\textbf{j}). 
The resonance of Si$_3$N$_4$ microresonator \#1(\#3) at 200.5(189.4) THz.
The intrinsic loss $\kappa_0/2\pi$ is 8.6(8.9) MHz and the external coupling rate $\kappa_\mathrm{ex}/2\pi$ is 4.7(8.4) MHz. 
Red circles are the measured data and the blue lines are the corresponding fits.
\textbf{h}(\textbf{k}). 
Measured intrinsic (blue dot) and external coupling rate (red dot) of Si$_3$N$_4$ microresonator \#1(\#3) over 20 THz (182.8 THz to 202.6 THz).
}
\label{Fig:S1}
\end{figure*}

Supplementary Fig.~\ref{Fig:S1}(d, e) shows the characterization of the DFB lasers \#1 and \#2.
As the current is tuned from 0 to 500 mA, the DFB laser \#1's output power increases to 159 mW with the laser threshold of 55 mA.
The wavelength can be tuned by 1 nm around 1548.882 nm at the temperature of 30 $^\circ$C. 
The current of DFB laser \#2 can be tuned up to 300 mA, allowing to output the maximum power of 83 mW.
The wavelength range is 0.6 nm around 1549.704 nm at the temperature of 25 $^\circ$C. 
The Si$_3$N$_4$ microresonators \#1 to \#3 feature normal group velocity dispersion (GVD), i.e. $D_2 < 0$, which enable the generation of platicons.
Si$_3$N$_4$ microresonators \#1 and \#3 are characterized as shown in Supplementary Fig.~\ref{Fig:S1}f--k, where Supplementary Fig.~\ref{Fig:S1}(f, i) shows the integrated dispersion $D_\mathrm{int}$, Supplementary Fig.~\ref{Fig:S1}(g, j) shows the typical resonance transmission, and Supplementary Fig.~\ref{Fig:S1}(h, k) shows the loss rate in wide bandwidth.
For Si$_3$N$_4$ microresonator \#1 at frequency $\nu = 200.5$ THz, the intrinsic loss $\kappa_{0}/2\pi$ and the external coupling rate $\kappa_\mathrm{ex}/2\pi$ are fitted as 8.6 MHz and 4.7 MHz respectively. 
The intrinsic quality factor is $Q_0 = 23 \times 10^6$.
For Si$_3$N$_4$ microresonator \#3 at frequency $\nu = 189.4$ THz, the intrinsic loss and external coupling rate are fitted as 8.9 MHz and 8.4 MHz respectively.
The intrinsic quality factor is $Q_0 = 21 \times 10^6$.
The comprehensive characterizations of the microresonators are performed by a home-developed vector spectrum analyzer\cite{Luo:24}.

\vspace{0.5cm}
\section{Universal Kerr-thermal dynamics in Setup \#2 and \#3}
\vspace{0.5cm}

\begin{figure*}[h!]
\renewcommand{\figurename}{Supplementary Figure}
\centering
\includegraphics{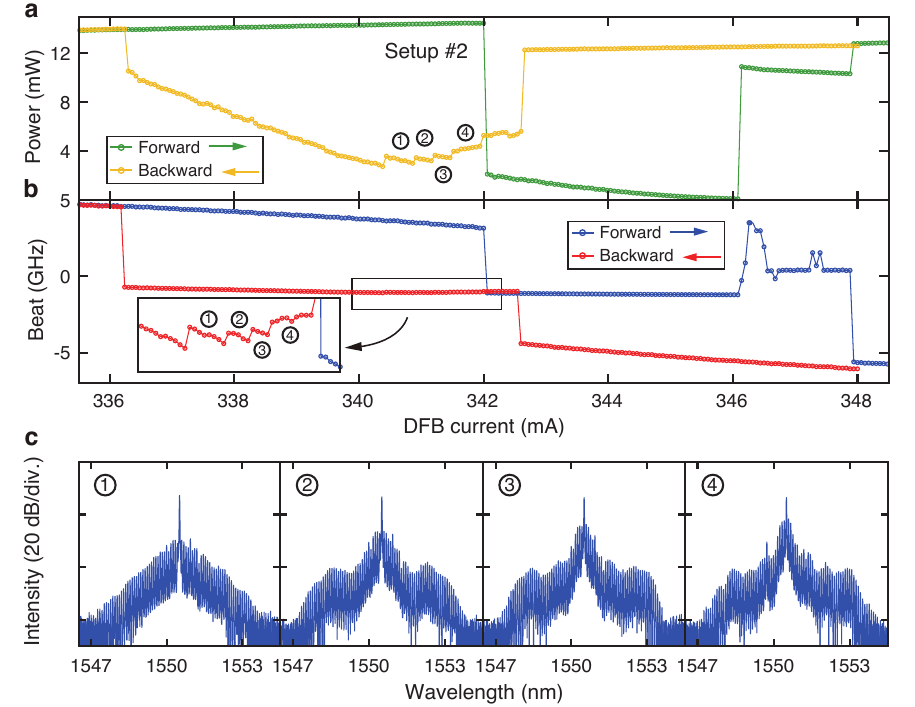}
\caption{
\textbf{SIL dynamics in the setup \#2}.
\textbf{a}. 
Measured microresonator output laser power in the forward (green) and backward (yellow) tuning processes.
The step features are distinctly presented in the backward tuning process.
\textbf{b}. 
Measured beat frequency between the microresonator output laser and a reference laser in the forward (blue) and backward (red) tuning processes. 
Inset is the zoom-in view on the step feature.
\textbf{c}. 
Measured optical spectra of the numbered steps in \textbf{b}. 
}
\label{Fig:S2}
\end{figure*}

Except in Setup \#1 (DFB laser \#1 $+$ Si$_3$N$_4$ microresonator \#1) in the main text, the Kerr-thermal dynamics are universally found in Setup \#2 (DFB laser \#2 $+$ Si$_3$N$_4$ microresonator \#2) and \#3 (DFB laser \#3 $+$ Si$_3$N$_4$ microresonator \#3), as shown in Supplementary Fig.~\ref{Fig:S2} and Supplementary Fig.~\ref{Fig:S3}.
Within the power and beat frequency traces, the platicon steps are observed in the backward tuning process.
Supplementary Fig.~\ref{Fig:S2}c(\ref{Fig:S3}c) displays the typical optical spectra at the numbered steps in Supplementary Fig.~\ref{Fig:S2}b(\ref{Fig:S3}b).

\begin{figure*}[t!]
\renewcommand{\figurename}{Supplementary Figure}
\centering
\includegraphics{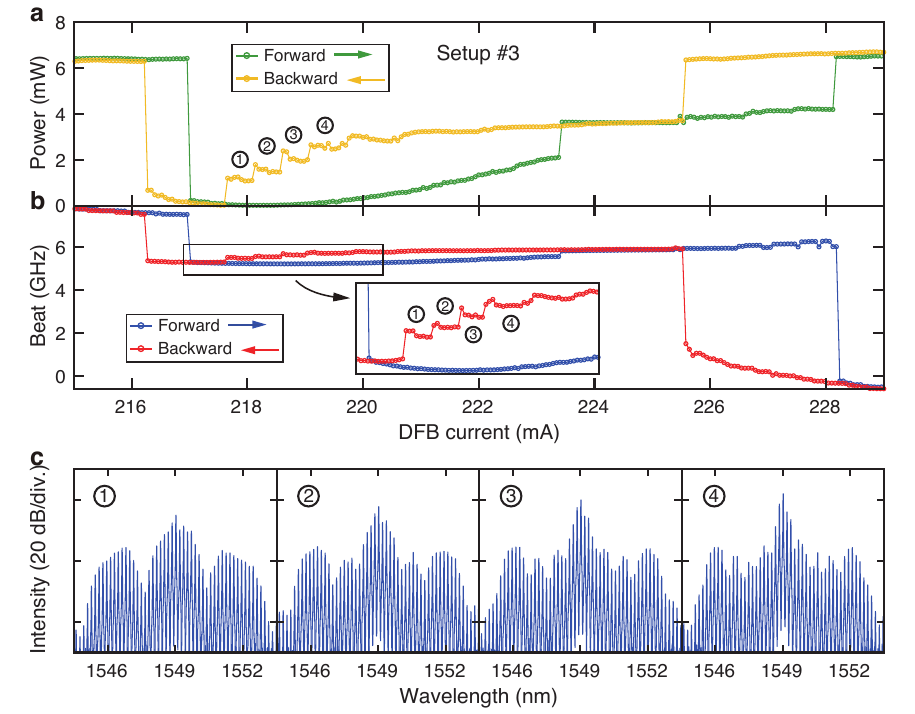}
\caption{
\textbf{SIL dynamics in the setup \#3}.
\textbf{a}. 
Measured microresonator output laser power in the forward (green) and backward (yellow) tuning processes.
\textbf{b}. 
Measured beat frequency between the microresonator output laser and the reference laser in the forward (blue) and backward (red) tuning processes. 
Inset is the zoom-in view on the step feature.
\textbf{c}. 
Measured optical spectra at the numbered steps in \textbf{b}. 
}
\label{Fig:S3}
\end{figure*}

\newpage
\section{Theory derivation and simulation parameters}
\vspace{0.5cm}

We begin from the laser rate equations that describe the dynamics of light field and carrier density in the FP laser diode, 
\begin{widetext}
\begin{align}
    \partial_tN&=\frac I{eV_\mathrm{L}}-\frac{\epsilon_0 n_\mathrm{L}^2}{\hbar\omega_\mathrm{L}}G(N)|E_\mathrm{L}|^2-\kappa_NN \label{eq:r1}\\
    \partial_tE_\mathrm{L}&=\left[\frac12G(N)+ i\delta\omega(N)-\frac12\alpha_0\right]E_\mathrm{L}+ i\text e^{ i(\delta\omega_{\mathrm{L}0}t+\omega_0t_\mathrm{s})}\frac{\sqrt{\kappa_{\mathrm{ex}}\tau_\mathrm{R} T_\mathrm{L} T_\mathrm{cL}}}{\tau_\mathrm{L}}\sqrt{\frac{n_0^2V_{\mathrm{eff}}\tau_\mathrm{L}}{2n_\mathrm{L}^2 V_\mathrm{L}\tau_\mathrm{R}}}E_0^-. \label{eq:r2}
\end{align}
\end{widetext}
Here $N$ is the carrier density. 
$I$ is the injection current. 
$V_\mathrm{L}$ is the effective mode volume of the gain area of the laser diode. 
$e$ is the carrier charge. 
$\epsilon_0$ is the vacuum permittivity. 
$n_\mathrm{L}$ is the refractive index of the gain area. 
$\hbar$ is the Planck constant. 
$\omega_\mathrm{L}$ is the resonance frequency of the cold laser cavity. 
$G(N)$ is the gain of the light intensity per unit time in the gain area, which depends on carrier density $N$. 
$E_\mathrm{L}$ is the amplitude of the light field in the gain area, which has omitted the phase factor $\text e^{- i\omega_\mathrm{L}t}$ and the transverse part. 
$\kappa_N$ is the dissipation rate of carrier density. 
$\delta\omega(N)$ is the frequency shift caused by the refractive index variation of the gain area induced by the carrier density. 
$\alpha_0$ is the loss factor of the light intensity. 

The last term in Eq.~\ref{eq:r2} describes the back-scattered light from the microresonator coupled into the laser cavity.
$\delta\omega_{\mathrm{L}0}=\omega_\mathrm{L}-\omega_0$ is the difference between $\omega_\mathrm{L}$ and the microresonator resonance $\omega_0$. 
$t_\mathrm{s}$ is the delay time between the microresonator and the laser cavity. 
$\kappa_{\mathrm{ex}}$ represents the external coupling rate of the microresonator.
$\tau_\mathrm{R}$ is the round-trip time of the microresonator. 
$T_\mathrm{L}$ is the transmission efficiency of the light intensity at the laser cavity's output surface. 
$T_{\mathrm{cL}}$ is the coupling efficiency of the light intensity into the laser cavity from outside. 
$\tau_\mathrm{L}$ is the round-trip time of the laser cavity. 
$n_0$ is the refractive index of the microresonator.
$V_\mathrm{eff}$ is the effective volume of the microresonator.
$n_\mathrm{L}$ is the refractive index of the active area.
$E_0^-$ is the light amplitude of the $0^{\mathrm{th}}$ mode of the back-scattered light in the microresonator. 

When the laser diode runs freely, i.e. without coupling the microresonator, the gain function $G(N)$ must equal to the loss factor $\alpha_0$ in order to reach a stable state. 
In the free-running case, we label the carrier density as $N_0$, the gain function as $G_0=G(N_0)$, and the frequency shift as $\delta\omega_0=\delta\omega(N_0)$. 
We assume that the carrier density $N$ does not change too far away from $N_0$ when the laser diode couples the microresonator.
This assumption is practically true, because the driving current $I$ does not change too much in the tuning process, as shown in Supplementary Fig.~\ref{Fig:S2} and Supplementary Fig.~\ref{Fig:S3}.
So, we can use linear approximation $G(N)\approx G_\mathrm{n}(N-N_0)+G_0$, and $\delta\omega(N)\approx\delta\omega_0+\alpha_\mathrm{g}G_\mathrm{n}(N-N_0)/2$ with $\alpha_\mathrm{g}$ being the linewidth enhancement factor.
After this approximation, we can rewrite the laser rate equations as below,
\begin{widetext}
\begin{align}
    \partial_tN&=\frac I{eV_\mathrm{L}}-\frac{\epsilon_0n_\mathrm{L}^2}{\hbar\omega_\mathrm{L}}[G_\mathrm{n}(N-N_0)+G_0]|E_\mathrm{L}|^2-\kappa_NN\\
    \partial_tE_\mathrm{L}&=\frac12(1+ i\alpha_\mathrm{g})G_\mathrm{n}(N-N_0)E_\mathrm{L}+i\mathrm{e}^{i(\delta\omega_\mathrm{L}t+\omega_0t_\mathrm{s})}\frac{\sqrt{\kappa_{\mathrm{ex}}\tau_\mathrm{R} T_\mathrm{L} T_\mathrm{cL}}}{\tau_\mathrm{L}}\sqrt{\frac{n_0^2V_{\mathrm{eff}}\tau_\mathrm{L}}{2n_\mathrm{L}^2 V_\mathrm{L}\tau_\mathrm{R}}}E_0^-.
\end{align}
\end{widetext}
Here we have rewritten the amplitude $E_\mathrm{L}=E_\mathrm{L}\text e^{i\delta\omega_0t}$, and $\delta\omega_{\mathrm{L}0}$ is replaced by $\delta\omega_\mathrm{L}=\delta\omega_{\mathrm{L}0}-\delta\omega_0$. 

Now we consider the equations describing the dynamics in the microresonator. 
We use the modified Lugiato-Lefever equation to describe the dynamics of the forward-propagating and back-scattered light field in the microresonator as below,
\begin{widetext}
\begin{align}
    \partial_tA_\mu^+&=-\left(\frac\kappa2+ i\sum_{n=2}^\infty\frac{D_n}{n!}\mu^n- iT_\mathrm{h}\right)A_\mu^+ + ig\left(\mathcal F[|A^+|^2A^+]_\mu+2A_\mu^+\sum_{\mu'}|A_{\mu'}^-|^2\right)+ i\beta A_\mu^-\nonumber\\
    &~~~~+ i\text e^{ i(-\delta\omega_\mathrm{L}t+\omega_\mathrm{L}t_\mathrm{s})}\sqrt{\frac{\kappa_{\mathrm{ex}}T_\mathrm{L}T_\mathrm{cR}}{\tau_\mathrm{L}}}A_\mathrm{L}\delta_{0\mu}\\
    \partial_tA_\mu^-&=-\left(\frac\kappa2+ i\sum_{n=2}^\infty\frac{D_n}{n!}\mu^n- iT_\mathrm{h}\right)A_\mu^- + ig\left(\mathcal F[|A^-|^2A^-]_\mu+2A_\mu^-\sum_{\mu'}|A_{\mu'}^+|^2\right)+ i\beta A_\mu^+\\
    \partial_tT_\mathrm{h}&=\frac1{\tau_\mathrm{h}}\left[K_T\sum_{\mu}(|A_\mu^+|^2+|A_\mu^-|^2)-T_\mathrm{h}\right].
\end{align}
\end{widetext}
Here $A_\mu^\pm=\mathcal F[A^\pm]_\mu=\int\text d\phi~A^\pm\text e^{i\mu\phi}$, $A^\pm$ is the forward-propagating/back-scattered laser amplitude which has been rescaled as
\begin{widetext}
\begin{align}
    A^\pm=\sqrt{\frac{\epsilon_0 n_0^2}{2\hbar\omega_0}V_{\mathrm{eff}}}E^\pm.
\end{align}
\end{widetext}
Also, $A_\mathrm{L}$ is the laser amplitude in the laser cavity and has a relation with $E_\mathrm{L}$ as 
\begin{widetext}
\begin{align}
    A_\mathrm{L}=\sqrt{\frac{\epsilon_0n_\mathrm{L}^2}{\hbar\omega_\mathrm{L}}V_\mathrm{L}}E_\mathrm{L}.
\end{align}
\end{widetext}
$\kappa=\kappa_0+\kappa_{\mathrm{ex}}$ is the loaded linewidth of the microresonator with $\kappa_0$ representing the intrinsic loss. 
$D_n$ is $n^\mathrm{th}$ order dispersion coefficient. 
$g=\hbar\omega_0^2cn_2/(n_0^2V_{\mathrm{eff}})$ is nonlinear coefficient, 
where $c$ is the light speed in vacuum,
$n_2$ is Kerr coefficient. 
$\beta$ is Raylaigh scattering coefficient. 
$T_{\mathrm{cR}}$ is the coupling efficiency of the laser to microresonator. 
$T_\mathrm{h}$ is the resonant frequency shift caused by thermal effect. 
$K_T$ is the thermal-induced frequency shift coefficient, 
and $\tau_\mathrm{h}$ is thermal relaxation time. 
Note that we have rescaled the laser amplitude $E_\mathrm{L}$ as $A_\mathrm{L}$, so the laser rate equations become
\begin{widetext}
\begin{align}
    \partial_tN&=\frac I{eV_\mathrm{L}}-\frac1{V_\mathrm{L}}[G_\mathrm{n}(N-N_0)+G_0]|A_\mathrm{L}|^2-\kappa_NN\\
    \partial_tA_\mathrm{L}&=\frac12(1+ i\alpha_\mathrm{g})G_\mathrm{n}(N-N_0)A_\mathrm{L}+i\text e^{i(\delta\omega_\mathrm{L}t+\omega_0t_\mathrm{s})}\sqrt{\frac{\kappa_{\mathrm{ex}} T_\mathrm{L} T_\mathrm{c}}{\tau_\mathrm{L}}}A_0^-.
\end{align}
\end{widetext}

By defining $\tau=\kappa t/2$, $a=\sqrt{2g/\kappa}A^+$, $b=\sqrt{2g/\kappa}A^-$, $a_{\text L}=\sqrt{2g/\kappa}A_\mathrm{L}$, $T=2T_\mathrm{h}/\kappa$, we normalize the equations above as 
\begin{widetext}
\begin{align}
\partial_{\tau}N &=J_\mathrm{n}-\tilde{\kappa}_{N}[g_\mathrm{n}(N-N_0)+g_0]|a_\mathrm{L}|^{2}-\kappa_{N}'N \\
\partial_\tau a_\mathrm{L} &=\frac{1}{2}\left (1+i\alpha_\mathrm{g}\right )g_\mathrm{n}(N-N_0)a_\mathrm{L}+i\mathrm{e}^{i(\alpha_\mathrm{L}\tau+\varphi)}\tilde{\kappa}_\mathrm{L}b_0 \\
\partial_\tau a_{\mu} &=-\left(1+i\sum_{n=2}^{\infty}\frac{d_{n}}{n!}\mu^{n}-iT\right)a_{\mu}+i\left(\mathcal{F}[|a|^{2}a]_{\mu}+2a_{\mu}\sum_{\mu^{\prime}}\left|b_{\mu^{\prime}}\right|^{2}\right)+i\tilde{\beta}b_{\mu}+i\mathrm{e}^{i(-\alpha_\mathrm{L}\tau+\varphi)}\tilde{\kappa}_\mathrm{R}a_\mathrm{L}\delta_{0\mu} \\
\partial_{\tau}b_{\mu} &=-\left(1+i\sum_{n=2}^{\infty}\frac{d_{n}}{n!}\mu^{n}-iT\right)b_{\mu}+i\left(\mathcal{F}[|b|^{2}b]_{\mu}+2b_{\mu}\sum_{\mu^{\prime}}\left|a_{\mu^{\prime}}\right|^{2}\right)+i\tilde{\beta}a_{\mu} \\
\partial_{\tau}T &=\left[k_{T}\sum_{\mu}\left(\left|a_{\mu}\right|^{2}+\left|b_{\mu}\right|^{2}\right)-T\right]/t_\mathrm{h} .
\end{align}
\end{widetext}
Here $J_\mathrm{n}=2I/(\kappa eV_{\text L})$, $\tilde\kappa_N=\kappa/(2gV_\mathrm{L})$, $g_\mathrm{n}=2G_\mathrm{n}/\kappa$, $g_0=2G_0/\kappa$, $\kappa_N'=2\kappa_N/\kappa$, $\alpha_{\text L}=2\delta\omega_\mathrm{L}/\kappa$, $\varphi=\omega_0t_\mathrm{s}$, $\tilde\kappa_{\mathrm{L}}$ are defined as
\begin{widetext}
\begin{align}
    \tilde\kappa_{\text L}=\frac2\kappa\sqrt{\frac{\kappa_{\mathrm{ex}} T_\mathrm{L} T_\mathrm{cL}}{\tau_\mathrm{L}}},
\end{align}
\end{widetext}
$d_n=2D_n/\kappa$, $\tilde\beta=2\beta/\kappa$, $\tilde\kappa_{\text R}$ is defined as
\begin{widetext}
\begin{align}
    \tilde\kappa_{\text R}=\frac2\kappa\sqrt{\frac{\kappa_\mathrm{ex}T_\mathrm{L}T_\mathrm{cR}}{\tau_{\mathrm{L}}}},
\end{align}
\end{widetext}
$k_T = K_T/g$, and $t_\mathrm{h} = \kappa \tau_\mathrm{h}/2$.

In order to make the parameters of the laser diode neater, we rescale $N$ and $a_{\text L}$ as $N\rightarrow N N_0$, $a_{\text L} \rightarrow \sqrt{(J_\mathrm{n}-\kappa_N'N_0)/(\tilde\kappa_N g_0)}a_{\text L}$, then the two laser equations become
\begin{widetext}
\begin{align}
    \partial_\tau N &= J_N-[(N-1)+\gamma]\frac{J_N-\kappa_N'}{\gamma}|a_{\text L}|^2-\kappa_N'N\\
    \partial_\tau a_{\text L} &= \frac12(1+i\alpha_\mathrm{g})\gamma_\mathrm{p}(N-1)a_{\text L}+i\text e^{i(\alpha_\mathrm{L}\tau+\varphi)}\tilde\kappa_\mathrm{L}\sqrt{\frac{\tilde\kappa_N g_\mathrm{n}\gamma}{J_N-\kappa_N'}}b_0.
\end{align}
\end{widetext}
Here $J_N = J_\mathrm{n}/N_0$, $\gamma = g_0/(g_\mathrm{n}N_0)$, $\gamma_\mathrm{p} = g_\mathrm{n}N_0$. Now we define $\gamma_\mathrm{e}$ as $\gamma_\mathrm{e} = (J_N-\kappa_N')/\gamma$, then the laser equations can be further transformed into
\begin{widetext}
\begin{align}
    \partial_\tau N &= \gamma_\mathrm{e}\left[ \gamma-F(N-1)-(N-1+\gamma)|a_{\text L}|^2 \right]\\
    \partial_\tau a_{\text L} &= \frac12(1+i\alpha_\mathrm{g})\gamma_\mathrm{p}(N-1)a_{\text L} + i\text e^{i(\alpha_\mathrm{L}\tau+\varphi)}\tilde\kappa_\mathrm{L}\sqrt{\frac{\tilde\kappa_Ng_\mathrm{n}}{\gamma_\mathrm{e}}}b_0
\end{align}
\end{widetext}
with $F = \kappa_N'/\gamma_\mathrm{e}$. At this point, we can see that $N$ represents the carrier density normalized to $N_{\text th}$, and $a_{\text L}$ represents the laser amplitude normalized to the equilibrium solution of the laser amplitude in free running which is $\sqrt{(J_\mathrm{n}-\kappa_N'N_0)/(\tilde\kappa_N g_0)}$. Now, comparing the new laser equations and the microresonator equations, we can define $f_\mathrm{e} = \tilde\kappa_\mathrm{R}\sqrt{\gamma_\mathrm{e}/(\tilde\kappa_Ng_\mathrm{n})}$ to represent the strength of the nonlinear effect in the microresonator and $\kappa_\mathrm{c} = \tilde\kappa_\mathrm{R}\tilde\kappa_\mathrm{L}$ to represent the coupling strength between the laser diode and the microresonator, then the ultimate equations become
\begin{widetext}
\begin{align}
    \partial_\tau N &= \gamma_\mathrm{e}\left[\gamma-F(N-1)-(N-1+\gamma)|a_{\text L}|^2\right] \label{eq:u1}\\
    \partial_\tau a_{\text L} &= \frac12\gamma_\mathrm{p}(1+i\alpha_\mathrm{g})(N-1)a_{\text L}+i\text e^{i(\alpha_\mathrm{L}\tau+\varphi)}\frac{\kappa_\mathrm{c}}{f_\mathrm{e}}b_0 \label{eq:u2}\\
    \partial_\tau a_{\mu} &=-\left(1+i\sum_{n=2}^{\infty}\frac{d_{n}}{n!}\mu^{n}-iT\right)a_{\mu}+i\left(\mathcal{F}[|a|^{2}a]_{\mu}+2a_{\mu}\sum_{\mu^{\prime}}\left|b_{\mu^{\prime}}\right|^{2}\right)+i\tilde{\beta}b_{\mu}+i\mathrm{e}^{i(-\alpha_\mathrm{L}\tau+\varphi)}f_\mathrm{e}a_\mathrm{L}\delta_{0\mu} \label{eq:u3}\\
    \partial_{\tau}b_{\mu} &=-\left(1+i\sum_{n=2}^{\infty}\frac{d_{n}}{n!}\mu^{n}-iT\right)b_{\mu}+i\left(\mathcal{F}[|b|^{2}b]_{\mu}+2b_{\mu}\sum_{\mu^{\prime}}\left|a_{\mu^{\prime}}\right|^{2}\right)+i\tilde{\beta}a_{\mu} \label{eq:u4}\\
    \partial_{\tau}T &=\left [ k_{T}\sum_{\mu}\left(\left|a_{\mu}\right|^{2}+\left|b_{\mu}\right|^{2}\right)-T\right ]/t_\mathrm{h} \label{eq:u5}.
\end{align}
\end{widetext}

The parameter values for the simulation results in the main text are listed in the Supplementary Table~\ref{Tab:1}.
The computer program for simulation usually requires several hours for a single run.

\begin{table*}[h!]
\renewcommand\tablename{Supplementary Table}
\renewcommand\arraystretch{1.5}
\center
\caption{Parameter values in simulation}
\begin{tabular}{|c|c|c|c|c|c|c|c|}
\hline
$I$~(mA) &  $V_\mathrm{L}$~(m$^3$) & $N_0$~(m$^{-3}$) & $\kappa_N$~(s$^{-1}$) & $G_\mathrm{n}$~(m$^3$/s) & $G_0$~(s$^{-1}$) & $\alpha_\mathrm{g}$ & $\alpha_\mathrm{L}$  \\
\hline
316.0 & 1.050$\times 10^{-14}$ & 1.000$\times 10^{24}$ & 3.507$\times 10^{7}$ & 9.524$\times 10^{-12}$ & 1.175$\times 10^{8}$ & 5.000 & 0  \\
\hline
$\tau_\mathrm{L}$~(s) & $\varphi$~($\pi$) & $T_\mathrm{L}$ & $T_\mathrm{cL}$ & $T_\mathrm{cR}$ & $\kappa/2\pi$~(MHz) & $\kappa_\mathrm{ex}/2\pi$~(MHz) &  $n_2$~(m$^2$/W) \\
\hline
3.803$\times 10^{-11}$ & 0.2000 & 0.7000 & 0.9000 & 0.3000 & 18.60 & 6.700 & 2.400$\times 10^{-19}$ \\
\hline
$n_0$ & $V_\mathrm{eff}$~(m$^3$) & $D_1/2\pi$~(GHz) & $D_2/2\pi$~(kHz) & $D_3/2\pi$~(Hz) & $\beta$ & $\tau_\mathrm{h}$~(s) & $K_T$~(rad/s/J)  \\
\hline
1.937 & 2.152$\times 10^{-14}$ & 10.70 & $-89.30$ & 11.50 & 0.6000 & $3.000\times 10^{-7}$ & 0.4170  \\
\hline
\end{tabular}
\label{Tab:1}
\end{table*}

\newpage
\section{Comparison with the conventional Lugiato-Lefever equation}
\vspace{0.5cm}

The Lugiato-Lefever equation (LLE) is widely used to model Kerr microcomb generation in optical microresonators. 
The modified LLE is written as \cite{Sun:25}
\begin{equation}
\frac{\partial a}{\partial\tau} = -a-i\mathcal F[\zeta_\mu a_\mu](\tau,\phi) + i|a|^2a+f,
\label{eq:LLE}
\end{equation}
where $a=\mathcal F[a_\mu](\tau,\phi)=\sum_\mu a_\mu(\tau)\text e^{i\mu\phi}$, $\zeta_\mu=-\xi+d_2\mu^2/2-\Delta_\mu$, $\xi=2(\omega_\mathrm{p}-\omega_0)/\kappa$ is the detuning (normalized to $\kappa/2$) between the pump laser frequency $\omega_\mathrm{p}$ and the microresonator resonance, 
$\Delta_\mu$ is the frequency shift of $\mu^\mathrm{th}$ mode normalized to $\kappa/2$,
$\phi$ is the azimuthal angle of the microresonator,
and $f=\sqrt{8g\kappa_\mathrm{ex}P_\mathrm{in}/\kappa^3}$ is the dimensionless pump amplitude with $P_\mathrm{in}$ being the on-chip pump laser power.

However, we emphasize that, although modified LLE can capture platicon steps \cite{Sun:25},it fails to describe the systems where SIL is involved. 
Compared with the comprehensive model developed in this work (Eq.~\ref{eq:u1}--\ref{eq:u5}), the main issues of LLE are listed below. 
\begin{itemize}

\item The LLE model only considers the forward-propagating laser in the microresonator, neglecting the laser dynamics in the DFB laser cavity (Eq.~(\ref{eq:u1}, \ref{eq:u2})) and the back-scattered laser dynamics in the microresonator (Eq.~\ref{eq:u4}).
As a result, the LLE model is intrinsically incapable to depict the process of SIL, such as frequency locking on the microresonator mode, laser frequency steps and the long platicon step induced by SIL.

\item The LLE model (Eq.~\ref{eq:LLE}) does not consider the universal thermal effects (Eq.~\ref{eq:u5}) in the microresonator, i.e. the photo-thermal effect and the thermal-optic effect. 
As a result, the LLE model (Eq.~\ref{eq:LLE}) fails to account for the laser frequency drift due to intracavity power build-up and the thermal effects.

\item Mode frequency shift $\Delta_\mu$ has to be intentionally introduced in the LLE model to excite the platicon state \cite{Lobanov:15}.
While the mode frequency shift is an artificial hypothesis in the LLE model, it is not essential in the model developed this work.
\end{itemize}

The comparison between these two models and the corresponding experimental observations are summarized in Supplementary Table \ref{Tab:2}.

\begin{table*}[h!]
\renewcommand\tablename{Supplementary Table}
\renewcommand\arraystretch{1.5}
\center
\caption{Models comparison}
\begin{threeparttable} 
\begin{tabular}{|c|c|c|c|}
\hline
\textbf{Comparison on the models} & LLE model & Model in this work & Experimental observation \\
\hline
Laser dynamics in the DFB cavity & $\times$ & $\bigcirc$ & - \\
\hline
Forward-propagating laser in the microresonator & $\bigcirc$ & $\bigcirc$  & - \\
\hline
Back-scattered laser in the microresonator & $\times$ & $\bigcirc$ & - \\
\hline
Thermal dynamics in the microresonator & $\times$ & $\bigcirc$ & - \\
\hline
Artificial mode frequency shift & $\bigcirc$ & $\times$ & - \\
\hline
\textbf{Comparison on the phenomena} &  &  & \\
\hline
DFB frequency locking to the microresonator & $\times$ & $\bigcirc$ & $\bigcirc$ \\
\hline
DFB laser power steps & $\bigcirc$ & $\bigcirc$ & $\bigcirc$ \\
\hline
DFB laser frequency steps & $\times$ & $\bigcirc$ & $\bigcirc$ \\
\hline
Platicon steps length &  1.5 MHz & $\approx$ 15 MHz & $\approx$ 60 MHz * \\
\hline
DFB laser frequency drift & $\times$ & $\bigcirc$ & $\bigcirc$ \\
\hline
\end{tabular}

\begin{tablenotes}
\item[] $\times$ not included, $\bigcirc$ included
\item[] * The step length in DFB current in Fig.~2\textbf{c} in the main text is about 0.2 mA and the frequency tuning coefficient is about $-0.3$ GHz/mA.
\end{tablenotes}
\end{threeparttable} 

\label{Tab:2}
\end{table*}

\newpage
\section{Analysis on noise quenching}
\vspace{0.5cm}

\begin{figure*}[h!]
\renewcommand{\figurename}{Supplementary Figure}
\centering
\includegraphics{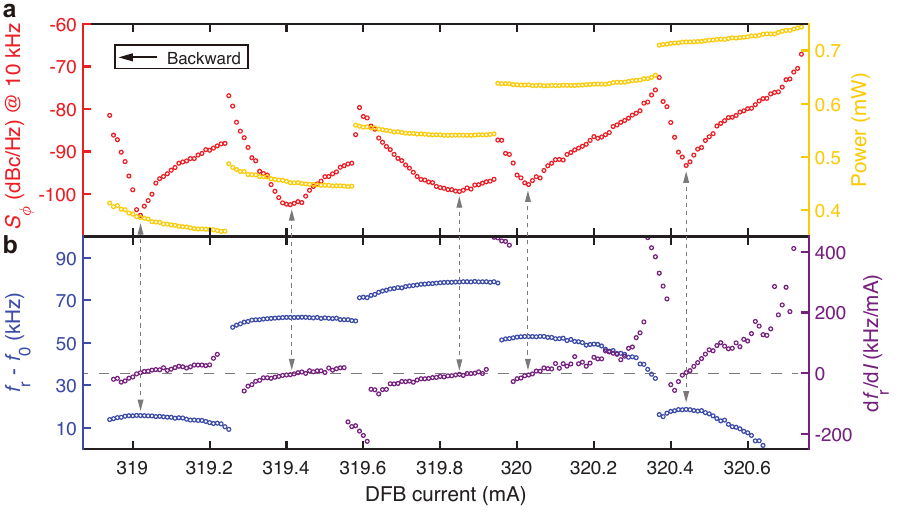}
\caption{
\textbf{Characterization of the noise quenching dynamics}.
\textbf{a}. 
Measured platicon optical power (yellow circles) and the microwave phase noise $S_\phi$ at 10 kHz Fourier offset frequency of the repetition rate $f_r$ (red circles).
\textbf{b}. 
Measured repetition rate $f_r$ (blue circles) and the derivative of $f_r$ to the DFB current (i.e. $\mathrm{d}f_r / \mathrm{d}I$, purple circles). 
$f_0 = 10.685450$ GHz.
Within each platicon state, the gray dashed arrows across \textbf{a} and \textbf{b} highlight the noise-quenching states, where the local minima of $S_\phi$ always coincide with the local maxima of $f_r$ and  $\mathrm{d}f_r / \mathrm{d}I = 0$ points.
}
\label{FigS:NQ}
\end{figure*}

Supplementary Fig.~\ref{FigS:NQ} shows five different platicon states accessed in the backward tuning as described in the main text (Setup \#1: DFB laser \#1 $+$ Si$_3$N$_4$ microresonator \#1). 
Upon photodetection of the platicon pulse stream, the PD outputs microwaves of carrier frequency corresponding to the platicon's repetition rate of $f_\mathrm{r}\approx10.7$ GHz.
We continuously tune the DFB current, and simultaneously monitor the platicon optical power (yellow data), the microwave phase noise (red data) and the repetition rate $f_\mathrm{r}$ (blue data). 
The switching dynamics among different platicon states endows discrete step features in the traces of the optical power. 
Within each platicon state, a local minimum of $S_\phi$ is found, providing phase noise quenching. 
Interestingly yet comprehensibly, such local minima of $S_\phi$ always coincide with local maxima of $f_\mathrm{r}$, where $f_\mathrm{r}$ has null dependence to the DFB current $I$, i.e. $\mathrm{d}f_\mathrm{r}/\mathrm{d}I=0$ as marked of the pruple data. 
It means that $f_\mathrm{r}$ is insensitive to current fluctuation and noise that are the main factors determining the laser frequency noise. 
Gray dashed arrows highlight the coincidence of local minima of $S_\phi$, local maxima of $f_\mathrm{r}$, and $\mathrm{d}f_\mathrm{r}/\mathrm{d}I=0$ points. 

\vspace{0.5cm}
\section{Universal noise quenching dynamics in different conditions}
\vspace{0.5cm}

\begin{figure*}[t!]
\renewcommand{\figurename}{Supplementary Figure}
\centering
\includegraphics{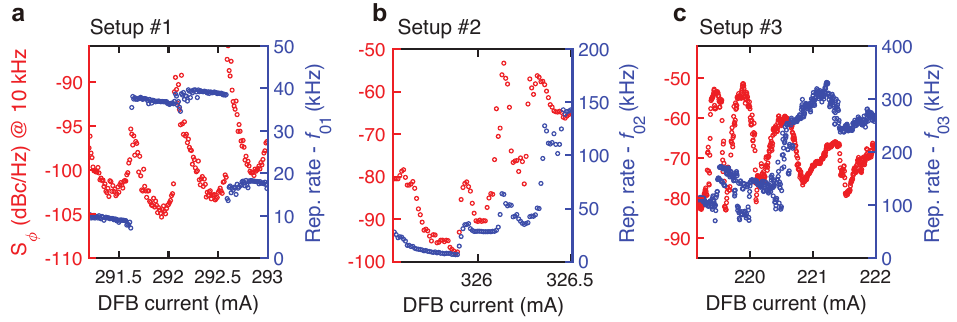}
\caption{
\textbf{Universal noise quenching dynamics}.
The red circles represent the phase noise of the platicon repetition rate at 10 kHz Fourier offset frequency. 
The blue circles represent the platicon repetition rate. 
Setup \#1 (DFB laser \#1 $+$ Si$_3$N$_4$ microresonator \#1) in \textbf{a} is the same setup as in the main text with Shijia Photons DFB laser and the 10.7-GHz-FSR Si$_3$N$_4$ microresonator. 
The microresonator resonance mode is different from that in the main text.
$f_{01} = 10.685340$ GHz. 
Setup \#2 (DFB laser \#2 $+$ Si$_3$N$_4$ microresonator \#2) in \textbf{b} contains the PhotonX Technology DFB laser and the 10.7-GHz-FSR Si$_3$N$_4$ microresonator. 
$f_{02} = 10.676660$ GHz. 
Setup \#3 (DFB laser \#3 $+$ Si$_3$N$_4$ microresonator \#3) in \textbf{c} contains the Shijia Photons DFB laser and the 21.3-GHz-FSR Si$_3$N$_4$ microresonator. 
$f_{03} = 21.371100$ GHz.
}
\label{FigS:UNQ}
\end{figure*}
The noise quenching phenomenon on palticon repetition rate can be found in different Si$_3$N$_4$ microresonator resonance modes and different setups, as shown in Supplementary Fig.~\ref{FigS:UNQ}. 
Supplementary Fig.~\ref{FigS:UNQ}a shows the noise quenching dynamics on the same setup (Setup \#1: DFB laser \#1 $+$ Si$_3$N$_4$ microresonator \#1) as in the main text but in a Si$_3$N$_4$ resonance mode with higher laser frequency (lower DFB current).
Supplementary Fig.~\ref{FigS:UNQ}(b, c) shows the noise quenching dynamics in Setup \#2 (DFB laser \#2 $+$ Si$_3$N$_4$ microresonator \#2) and Setup \#3 (DFB laser \#3 $+$ Si$_3$N$_4$ microresonator \#3).
All the Si$_3$N$_4$ microresonators have normal GVD, permitting platicon generation, as characterized in Supplementary Fig.~\ref{Fig:S1}.
The repetition rates of the platicons have local maximum or minimum values, exhibiting null dependence on the DFB current.
At the null dependence pionts, the phase noises are quenched to local minimum values.



\vspace{1cm}
\section*{Supplementary References}
\bigskip
\bibliographystyle{apsrev4-1}
\bibliography{bibliography}